# Hardware Acceleration of Kolmogorov–Arnold Network (KAN) in Large-Scale Systems

Wei-Hsing Huang, Jianwei Jia, Yuyao Kong, Faaiq Waqar, Tai-Hao Wen, Meng-Fan Chang, *Fellow, IEEE*, Shimeng Yu, *Fellow, IEEE*

**Abstract**—Recent developments have introduced Kolmogorov-Arnold Networks (KAN), an innovative architectural paradigm capable of replicating conventional deep neural network (DNN) capabilities while utilizing significantly reduced parameter counts through the employment of parameterized B-spline functions incorporating trainable coefficients. Nevertheless, the B-spline functional components inherent to KAN architectures introduce distinct hardware acceleration complexities. While B-spline function evaluation can be accomplished through look-up table (LUT) implementations that directly encode functional mappings, thus minimizing computational overhead, such approaches continue to demand considerable circuit infrastructure, including LUTs, multiplexers, decoders, and associated components. This work presents an algorithm-hardware co-design approach for KAN acceleration. At the algorithmic level, techniques include Alignment-Symmetry and PowerGap KAN hardware aware quantization, KAN sparsity aware mapping strategy, and circuit-level techniques include N:1 Time Modulation Dynamic Voltage input generator with analog-compute-in-memory (ACIM) circuits. Furthermore, this work conducts comprehensive evaluations on large-scale KAN networks to validate the proposed methodologies. Non-ideality factors, including partial sum deviations arising from process variations, have been evaluated with the statistics measured from the TSMC 22nm RRAM-ACIM prototype chips. Utilizing optimally determined KAN hyperparameters in conjunction with circuit optimizations fabricated at the 22nm technology node, despite the parameter count for large-scale tasks in this work increasing by 500K× to 807K× compared to tiny-scale tasks in previous work, the area overhead increases by only 28K× to 41K×, with power consumption rising by merely 51× to 94×, while accuracy degradation remains minimal at 0.11% to 0.23%, thereby demonstrating the scaling potential of our proposed architecture.

**Keywords**—Kolmogorov-Arnold Networks (KAN), Quantization, Compute-in-Memory (CIM), Resistive Random Access Memory (RRAM), Algorithm-Hardware Co-Design

This work was supported in part by the PRISM, one of the SRC/DARPA JUMP 2.0 centers.

Wei-Hsing Huang, Jianwei Jia, Yuyao Kong, Faaiq Waqar, and Shimeng Yu are with the School of Electrical and Computer Engineering, Georgia Institute of Technology, Atlanta, GA 30332 USA (Corresponding author. E-mail: shimeng.yu@ece.gatech.edu).
Wei-Hsing Huang and Jianwei Jia are co-first authors.
Tai-Hao Wen is the Department of Electrical Engineering, National Tsing Hua University (NTHU), Hsinchu 30013, Taiwan.
Meng-Fan Chang is with the Department of Electrical Engineering, National Tsing Hua University (NTHU), Hsinchu 30013, Taiwan, and also with Taiwan Semiconductor Manufacturing Company (TSMC), Hsinchu 30075, Taiwan.

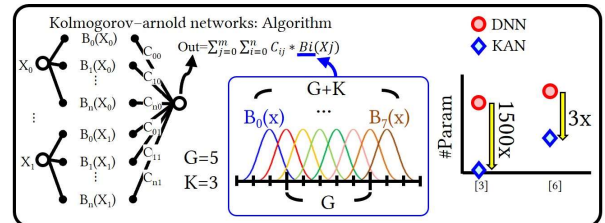

**Fig. 1.** Introduction of KAN and its potential for parameter reduction.

## 1 INTRODUCTION

Contemporary deep neural network (DNN) models characterized by their ever-escalating parameter counts present significant impediments to edge device deployment, substantially constraining the implementation of privacy-preserving, real-time detection capabilities and severely limiting the feasibility of resource-constrained edge computing applications. Traditional deep neural networks, including convolutional neural network (CNN) designs, large language model (LLM) frameworks, and various other architectural paradigms, conventionally implement fixed, predetermined activation functions coupled with trainable weight parameters [1][2]. The recently emerged Kolmogorov-Arnold Networks (KAN) [3], which drew inspiration from the mathematical foundations of the Kolmogorov-Arnold representation theorem [4][5], fundamentally reimagines the traditional multi-layer perceptron (MLP) architecture by replacing conventional linear weight matrices with parameterized B-spline functions B(X) while simultaneously implementing trainable, learnable activation functions positioned on the network edges rather than nodes. According to recent studies, KANs have demonstrated the ability to achieve comparable or superior accuracy with networks that are smaller in parameter count compared to traditional MLPs [3][23]. This innovative architectural paradigm has been empirically validated, demonstrating improvements not only in performance but also in interpretability, as comprehensively illustrated in Fig. 1 [3][6]. KAN architectures demonstrate considerable potential and promise to successfully replace conventional traditional DNN backbones that utilize fixed activation functions combined with learnable weights in increasingly complex DNN models, potentially enabling substantial reduction in the overall size of large-scale models, including computationally intensive LLMs and recommendation models as MLPs are widely used as building blocks, and thereby facilitating their practical deployment on edge devices.

However, despite these advantages, KAN operation fundamentally requires computationally intensive B-spline function computation and evaluation. While classical mathematical definitions involving recursive computational methods [7] can theoretically evaluate B-spline functions with mathematical precision, the computational requirements and processing demands increase dramatically and significantly with progressively higher-order k values. An alternative approach more suitable for edge-friendly implementation employs pre-computed lookup tables (LUTs) for direct and immediate B-spline function mapping, substantially simplifying the hardware implementation complexity and dramatically reducing the overall computational demands. Despite these considerable advantages and benefits, this implementation approach still necessitates and requires significant circuit resources, including extensive LUTs, multiplexers (MUXs), decoders, and associated control logic, as comprehensively depicted and shown in Fig. 2.

Furthermore, KAN architectures, similar to traditional MLPs, inherently involve extensive and computationally intensive parallel multiply-accumulate (MAC) operations throughout their computational pipeline. The well-documented von Neumann bottleneck present in conventional computing architectures leads to substantial inefficiencies and performance degradation. This bottleneck results from the physical separation between memory and processing units, causing significant data movement overhead that can dominate the overall energy consumption in data-intensive applications. Compute-in-Memory (CIM) [8], representing an emerging non-von Neumann architectural paradigm, directly addresses and mitigates this fundamental issue. CIM variants include digital-CIM (DCIM) [9], SRAM analog-CIM (ACIM) [10][11], and RRAM-ACIM [12][13], etc. While DCIM and SRAM-ACIM architectures offer comparatively higher computational accuracy than RRAM-ACIM implementations, the inherently large SRAM cell sizes substantially limit achievable on-chip storage capacity, and their characteristically high standby power consumption proves particularly undesirable and problematic for battery-constrained edge devices. Therefore, this paper comprehensively examines and investigates RRAM-ACIM acceleration techniques for KAN architectures. However, it should be explicitly noted that the proposed algorithm-level optimizations and enhancements are fundamentally hardware-agnostic and can be applied across different implementation platforms.

Despite RRAM-ACIM's numerous advantages for practical edge deployment scenarios, including remarkably low standby power consumption and high integration density, it encounters several critical challenges as comprehensively illustrated in Fig. 2. The process-temperature-voltage (PVT) variations are key concerns for ACIM. The progressively increasing array sizes combined with aggressive technology scaling substantially exacerbate IR-drop phenomena [14], primarily due to increased parasitic bit line resistance, thereby hampering and degrading inference accuracy performance. Moreover, current mainstream CIM input methodologies include binary input schemes with multi-cycle operation, multi-voltage level input approaches within single cycle operation, or time delay mechanisms with multi-bit input encoding. Binary input methodology offers superior accuracy but consequently increases operational latency and fundamentally limits achievable TOPS/W metrics. While multi-voltage level input approaches successfully achieve favorable and attractive TOPS/W ratios combined with low operational latency, the inherently constrained VDD range renders these inputs particularly susceptible to noise interference, thereby substantially compromising computational accuracy. Time-delay multi-bit input schemes offer enhanced noise resilience capabilities but at the considerable cost of increased operational latency and reduced throughput.

In this work, we systematically aim to co-optimize both algorithmic and circuit-level implementations, strategically reducing area, power consumption, and operational latency while simultaneously increasing the computational accuracy of ACIM-based computation specifically tailored for KAN architectures. This paper presents a journal extension of our previous conference presentation [24]. We have augmented all four schemes with comprehensive implementation details and introduced two additional algorithms to elaborate on the framework's technical specifications. Most importantly, our prior work [24] limited its evaluation to small-scale models (with 279~2232 B) within the context of knot theory, this work employs large-scale KAN models (39 MB ~ 63 MB) for a recommendation system [23] to provide a scaling aspect of our proposed methodology. The evaluation in Section 4 is completely redone for the large-scale KAN models.

## 2 BACKGROUND

### 2.1 KAN: Kolmogorov-Arnold Networks and acceleration

In contrast to MLPs, whose theoretical foundation rests on the universal approximation theorem, KAN architectures are grounded in the Kolmogorov-Arnold representation theorem. The mathematical formulation of individual KAN layers follows equations (1)-(2), wherein the b(x) component, initially implemented through SiLU activation functions, incorporates residual connectivity into the architecture.

$$\phi(x) = w_b b(x) + w_s \text{spline}(x) \qquad (1)$$

$$\text{spline}(x) = \sum_i c_i \, B_i(x) \qquad (2)$$

$$\phi(x) = w_b b(x) + \sum_i c_i' \, B_i(x) \qquad (3)$$

Fig. 1 depicts the relationship between the grid size parameter G and the B-spline order K, where the aggregate count of $B_i(x)$ functions equals K+G. For the presented configuration with K=3 and G=5, this work substitutes ReLU activation functions for the original SiLU implementation, achieving enhanced hardware efficiency while maintaining computational accuracy. Given that the $w_b b(x)$ term can be efficiently accelerated through conventional RRAM-ACIM architectures, the primary optimization target becomes the $w_s$spline(x) computation. The implementation strategy involves the multiplication of $w_s$ with $c_i$ to generate $c_i'$, followed by 8-bit quantization, yielding the modified formulation expressed in equation (3). This configuration enables $c_i'$ storage within the RRAM-ACIM array, where $B_i(x)$ values are delivered through word lines (WL) to facilitate parallelized multiply-accumulate computations. LUT implementations for $B_i(x)$ computation demonstrate superior suitability for edge deployment scenarios relative to recursive B-spline evaluation methodologies. Furthermore, the uniform nodal distribution inherent in KAN architectures ensures that B(X) functional representations remain consistent across varying knot grid intervals, thus enabling the feasibility of shared LUT architectures for multiple B(X) instantiations. Nevertheless, as illustrated in Fig. 2, existing quantization approaches [15][16] introduce misalignment between knot grid and quantization grid structures, generating unique input-output mapping relationships for individual $B_i(x)$ functions. Consequently, hardware implementations require dedicated LUT, multiplexer, and decoder resources for each $B_i(x)$ component at the edge, leading to substantial energy consumption and silicon area penalties. While the adoption of fixed, non-programmable LUT architectures offers a potential mitigation strategy through reduced circuit footprint, such implementations sacrifice architectural flexibility, particularly the capability to dynamically modulate B(X) computational precision according to application-specific constraints. Additionally, the proliferation of $B_i(x)$ functional units corresponding to increased K and G parameters fundamentally limits the deployment viability of traditional quantization strategies for sophisticated KAN architectures characterized by elevated K and G values within resource-constrained edge environments. The Alignment-Symmetry and PowerGap hardware-aware quantization methodology, detailed in Section 3.1, is developed to overcome these limitations. Prior work [25] explored the implementation of Piece-Wise Linear (PWL) approximations as alternatives to B-spline functions through one-hot and thermometer encoding schemes. However, as B-spline complexity increases, PWL requires correspondingly higher precision, and the use of one-hot or thermometer encoding paradigms incurs substantial memory overhead, thereby limiting their practical applicability, particularly in large-scale and more complex tasks. Prior work [26] employed analog resistor networks utilizing series and parallel configurations for B-spline function computation. However, this approach demands exceptionally high precision in resistance values, presenting significant limitations in high-precision application scenarios, particularly for 8-bit implementations. Prior work [27] implemented KAN nonlinearities via stochastic computing—using a Segmented Multi-Dimensional Multi-Driving Finite State Machine (SMM-FSM) driven by pseudo-random bitstreams and crossbar-routed interconnects—but, compared with our deterministic one-cycle shared-LUT B-spline ACIM co-design, it suffers from long bit-stream latency and substantial overhead dominated by parameter memory and random number generation. Moreover, these prior works [25,26,27] have only been applied to small-scale tasks, whereas this work conducts a comprehensive evaluation on large-scale tasks.

## 2.2 Compute-in-memory

CIM architectures utilize various embedded memory technologies, including SRAM, eDRAM, and emerging non-volatile memories such as RRAM. These memory architectures exhibit distinctive operational characteristics and confront unique implementation constraints. SRAM-based CIM implementations, while benefiting from advanced process node availability and superior access latency characteristics, are constrained by substantial leakage current, rendering them suboptimal for deployment in energy-constrained edge computing environments where standby power minimization remains critical. Non-volatile memory technologies present attractive characteristics for edge deployment scenarios, particularly through reduced quiescent power dissipation and enhanced storage density at mature process nodes, translating to favorable cost effectiveness. Nevertheless, voltage degradation along bit lines in RRAM-ACIM configurations introduces inference precision limitations. Additionally, traditional CIM implementations predominantly utilize either voltage-domain or temporal-domain modulation schemes for encoding multi-bit word line inputs within single operational cycles. These modulation approaches demonstrate vulnerability to on-chip interference, resulting in compromised computational accuracy as illustrated in Fig. 2. The N:1 Time Modulation Dynamic Voltage Input Generator and KAN sparsity-aware weight mapping strategies, elaborated in Sections 3.2 and 3.3 respectively, provide systematic solutions to these implementation challenges within RRAM-ACIM frameworks.

## 3 PROPOSED TOP-DOWN HW-SW CO-OPTIMIZATION

### 3.1 *Alignment-Symmetry and PowerGap KAN hardware aware quantization for B(X)*

The proposed Alignment-Symmetry and PowerGap KAN hardware aware quantization (ASP-KAN-HAQ) is developed to minimize computational overhead and energy consumption associated with B(X) function evaluation as defined in Equation (3). This approach encompasses a two-stage optimization strategy:
- Phase one: Alignment-Symmetry for suppressing the needs of programmable LUT by enabling zero offset between knot grid and quantization grid.
- Phase two: PowerGap for reducing decoder and multiplexer complexity by constraining knot grid intervals to power-of-two magnitudes.

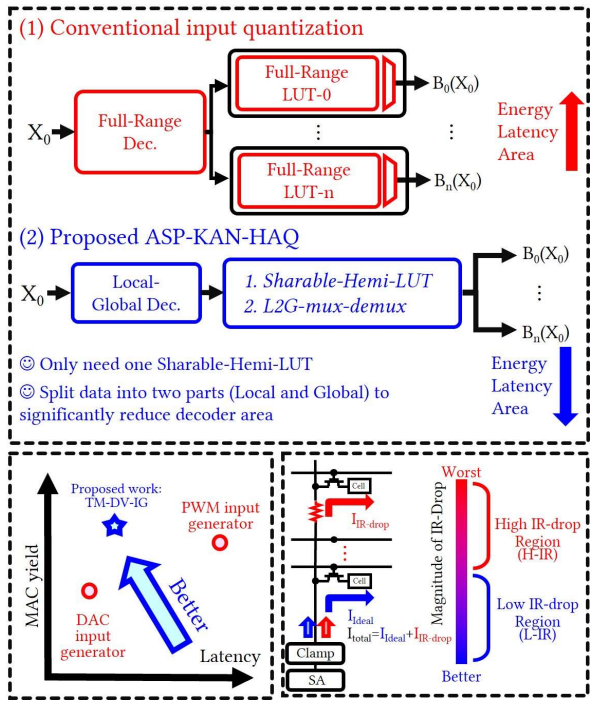

**Fig. 2.** Challenges hindering low-power and high accuracy edge AI applications.

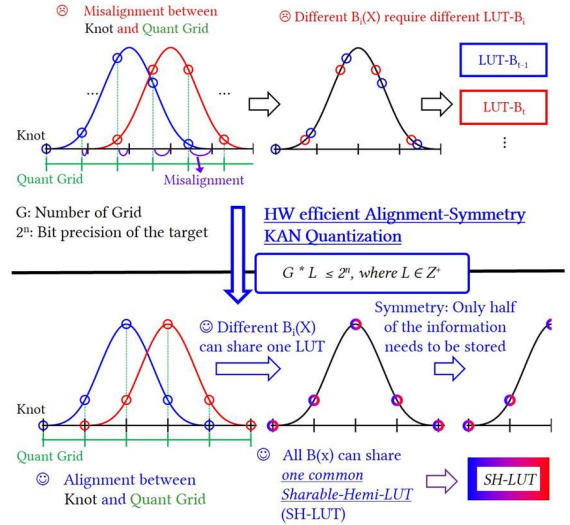

**Fig. 3.** HW efficient Alignment-Symmetry KAN Quantization for LUTs optimization.

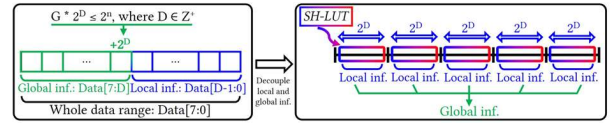

**Fig. 4.** HW efficient PowerGap KAN Quantization for MUXs and Decoders optimization.

The subsequent analytical framework examines a representative configuration with parameters K=3 and G=5, yielding eight basis functions $B_0(x)$ through $B_7(x)$ per input channel. The implementation assumes 8-bit input quantization with values spanning the integer range [0, 255]. Notably, the ASP-KAN-HAQ framework maintains extensibility to arbitrary parameter configurations of K and G, variable precision specifications, and architectural layers incorporating negative-valued inputs.

A. Phase One: Alignment-Symmetric

The first phase, designated as Alignment-Symmetric, derives from empirical observations presented in Fig. 3. The misalignment between knot grid and quantization grid prevents the utilization of a unified LUT across multiple B(X) functional instances, despite potential data translation from disparate knot grid intervals into a common interval space. This limitation is resolved through the ASP-KAN-HAQ framework, which establishes precise alignment between knot and quantization grid structures for individual B(X) functions. This alignment is achieved by imposing a constraint whereby the quantization grid dimensions constitute integer multiples of the corresponding knot grid parameters, formulated as:

$$G * L \leq 2^n, \text{ where } L \in Z+ \quad (4)$$

In Equation (4), the parameters G and n represent the knot count and the system's maximum bit-width specification, respectively. The value of L that satisfies Equation (4) constrains the data range to the interval [0, G*L-1]. Values of L adhering to this integer multiple relationship eliminate positional discrepancies between knot and quantization grids, thereby facilitating the deployment of a unified LUT architecture across all B(X) functional components. Additionally, this constraint induces symmetrical properties within the quantized B(X) representations, which permits a 50% reduction in shared LUT memory requirements. This optimized architectural configuration is designated as a Sharable-Hemi LUT (SH-LUT).

Following the Alignment-Symmetric phase, a direct implementation strategy for value routing from the SH-LUT to respective $B_0(x)$ through $B_7(x)$ functions with reduced hardware complexity utilizes eight 2L-to-1 transmission gate multiplexers (TG-MUXs) alongside an 8-bit optimized decoder architecture. Nevertheless, this configuration continues to exhibit substantial silicon area requirements and elevated power dissipation characteristics.

B. Phase Two: PowerGap

The second phase, designated as PowerGap, is developed to minimize transmission gate MUXs (TG-MUX) and decoder overhead following the Alignment-Symmetric optimization. Through constraining knot grid intervals to power-of-two values, this approach decouples local from global information domains, substantially reducing decoder and TG-MUX area requirements as illustrated in Fig. 4, with the mathematical representation expressed as:

$$G * 2^D \leq 2^n, \text{ where } D \in Z+ \quad (5)$$

Within the KAN architecture, information contained in individual knot grids is characterized as local information, whereas the mapping between distinct grid intervals and their corresponding B(X) functions constitutes global information. This distinction enables substantial reductions in hardware resource utilization.

The hardware requirements are significantly reduced:

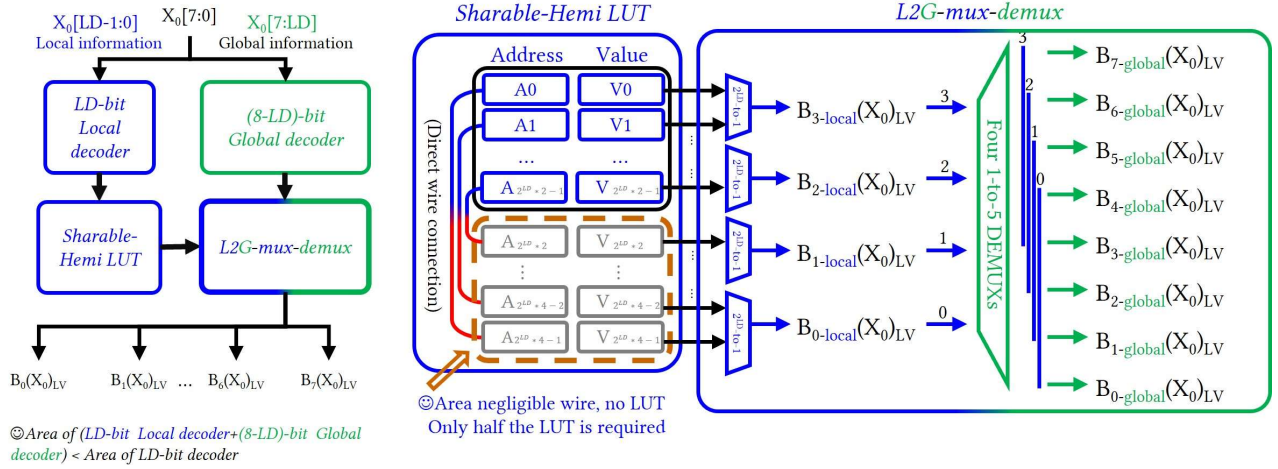

**Fig. 5.** Hardware architecture with Alignment-Symmetry and PowerGap KAN hardware aware quantization.

1. TG-MUXs: from original eight 2L-to-1 TG-MUXs to optimized four L-to-1 TG-MUXs and four 1-to-5 TG-DEMUXs.
2. Decoders: from original one 8-bit decoder to optimized one (8-D)-bit decoder and one D-bit decoder.

Since decoder area scales exponentially with bit-width specifications, the silicon footprint of a single 8-bit decoder substantially surpasses the combined area of an (8-D)-bit decoder and a D-bit decoder. Consequently, parameter values that simultaneously satisfy the constraints imposed by Equations (4) and (5) achieve optimal area reduction across LUT, decoder, and TG-MUX components, as mathematically formulated by:

$$G * 2^{LD} \leq 2^n, \text{where } LD \in Z+ \qquad (6)$$

We designate this optimal value as LD and this value constrains the data range to the interval $[0, G*2^{LD}-1]$. Fig. 5 depicts the hardware architecture and dataflow for B(X) lookup operations following ASP-KAN-HAQ optimization. Please note that Fig. 5 presents a high-level architectural diagram. As shown in Fig. 6, during actual implementation, when the Quant Grid is even numbered, the entire LUT can be partitioned into two halves for mutual sharing. When the Quant Grid is odd numbered, all LUTs remain shareable except for the central LUT. Since only a single LUT cannot be shared, the additional overhead incurred by odd numbered Quant Grids is negligible. Fig. 7 shows the efficient lookup process wherein multiple Xi values share a single SH-LUT, facilitating the transfer of corresponding LUT values ($B_{0\sim7\text{-global}} (Xi)_{LV}$) from local to global scope, which are subsequently propagated to the input generator.

### 3.2 N:1 Time Modulation Dynamic Voltage Input Generator for ACIM for $\sum c_i' B_i(X)$

In conventional CIM architectures, multi-bit WL input methods are typically realized through either pure voltage modulation [18][19] or pure pulse-width modulation (PWM) [20][21]. Voltage modulation encodes WL weights directly into different amplitude levels, but the voltage interval between adjacent levels becomes narrow as the bit resolution increases, which inevitably amplifies the impact of device variation, supply noise, and nonlinearity of the MOSFET transfer curve, leading to poor robustness. In contrast, PWM maps information into temporal width differences, which provides better resilience against small analog variations; however, distinguishing multiple bit levels requires long pulse widths, thereby extending the MAC cycle and severely degrading throughput. To overcome these limitations, we develop an N:1 Time-Modulated Dynamic Voltage Input Generator (TM-DV-IG), which maps LUT values B(X) into multi-level WL signals by simultaneously exploiting both the voltage and time domains. The WL input is encoded as a dynamic voltage pulse whose amplitude and width jointly determine the BL charging process, thus distributing information across two orthogonal domains. For a single RRAM cell, the BL current is proportional to the WL voltage, i.e., $I[x] \propto f(V[x]), x \in [0, 2^N - 1]$, where $f$ denotes the MOSFET transfer function; when this current flows for a duration $t = W[x]$, the accumulated charge becomes $Q = I[x] \cdot W[x]$. By carefully designing the DAC output voltages $V[2^N - 1:0]$ such that the current ratios satisfy $I[0]: I[1]: I[2] ... : I[2^N - 1] = 0: 1: 2: ... : 2^N - 1$, the unit interval of charge is defined as $W_{P1} \cdot I[1]$, which enables a linear distribution of charge values Q across all bit combinations. Compared with pure voltage input methods, this hybrid approach enhances tolerance to noise and device variation, while compared with pure PWM schemes, it avoids long pulses and maintains high operation speed (>100 MHz), thereby enabling multi-bit MAC execution within a single clock cycle.

The TM-DV-IG is composed of five major components: a Delay Chain, a Pulse Modulation and Timing Control Module (PM-TCM), an N-bit DAC, a Transmission-Gate Multiplexer (TG-MUX), and a Buffer Array, as illustrated in Fig. 8. The PM-TCL not only generates the control signals for buffer array voltage switching but also works with the delay chain to generate ratioed pulses $W_{P1}$, $W_{PN}$, and $W_{P(N+1)}$ with timing proportions of $1: 2^N: 2^N+1$. This design eliminates the need for counter-based digital logic, thereby saving area. The N-bit DAC

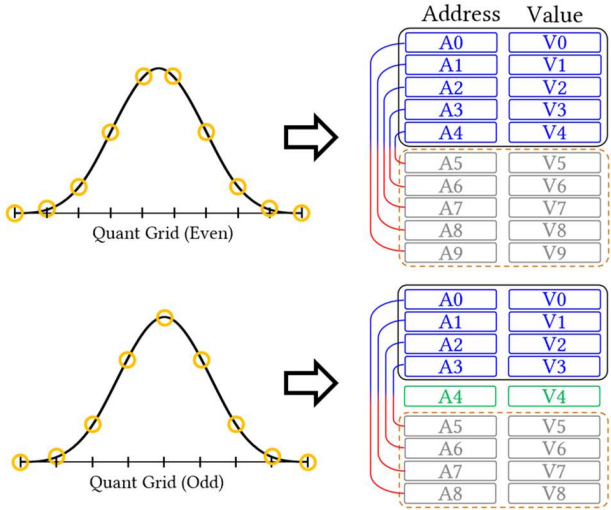

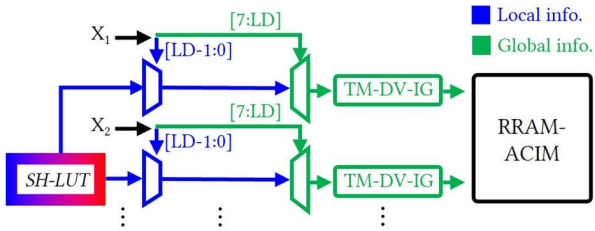

**Fig. 6.** The hardware architecture with efficient LUT retrieval process.

**Fig. 7.** The hardware architecture with efficient LUT retrieval process.

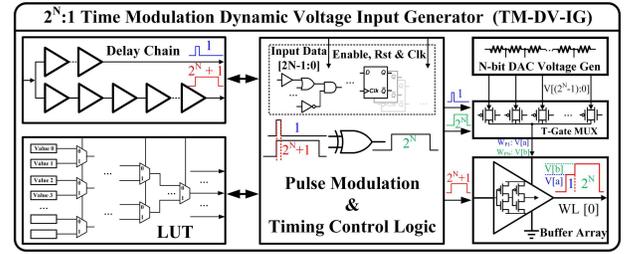

**Fig. 8.** N:1 Time Modulation Dynamic Voltage input generator for ACIM.

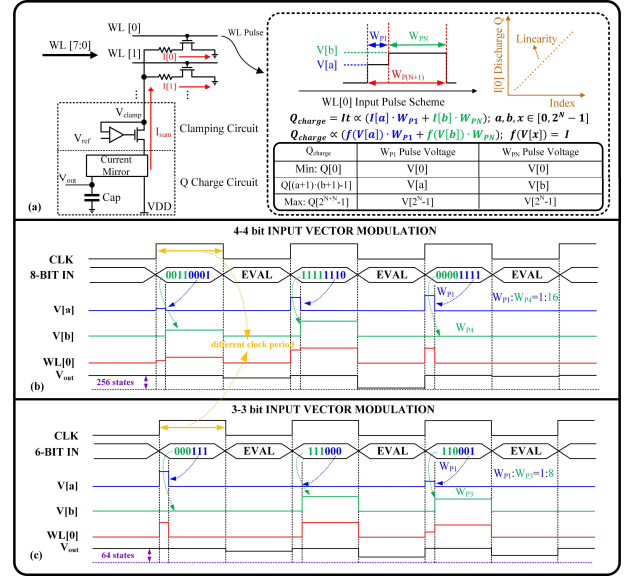

**Fig. 9.** (a) BL linear Q value generation theory (b) 3-3 bit input vector scheme for high accuracy application; (c) 4-4 bit input vector schemes for high speed application

produces $2^N$ distinct fixed voltage levels, which are selectively connected to the buffer array through the TG-MUX under control of PM-TCM pulses. The PM-TCM arranges these pulses to drive the TG-MUX switches so that different supply voltages V[$2^N - 1: 0$] are dynamically assigned to the buffer array according to the required LUT mapping. Meanwhile, the $2^N+1$ ratio pulse is applied directly to the buffer array, whose outputs are connected to the WLs. During read operation, the PM-TCM first receives the 2N-bit input vector, consults the LUT to determine the required pulse-voltage combination, and then generates the corresponding WL input waveform. On the BL side as shown in Fig.9 (a), a clamping circuit holds the BL voltage at a reference level $V_{clamp}$. In idle mode, precharge transistors initialize the capacitor to VDD; in read mode, different combinations of WL voltage and pulse duration discharge the capacitor with distinct currents, producing unique charge levels Q. These charge differences are then sensed by the sense amplifier (SA), realizing a linear and robust mapping of digital inputs to analog charge values. Importantly, the TM-DV-IG supports circuit reusability across multiple WLs, allowing most peripheral blocks to be shared and thereby minimizing area overhead. Furthermore, by adjusting the design parameter N, the architecture can be flexibly optimized for different operating modes. In the high-performance mode (TD-P), as illustrated in Fig. 9(b) with N=4, throughput is prioritized: during the positive half clock cycle, an 8-bit input vector is applied to the TM-DV-IG, where the lower 4 bits control the $W_{P1}$ pulse voltage modulation (V[a]) and the upper 4 bits control the $W_{P3}$ pulse voltage modulation (V[b]). The resulting WL[0] input is the combined dynamic voltage pulse of V[a] and V[b], producing an effective output voltage $V_{out}$ on the sampling capacitor. This configuration yields $8 \times 8 = 64$ distinct voltage states, enabling dense encoding. In contrast, in the high-accuracy mode (TD-A), as shown in Fig. 9(c) with N=3, the design provides finer charge resolution for precision-critical tasks. This adaptability, together with the dual-domain encoding principle, makes TM-DV-IG a scalable and efficient solution for next-generation multi-bit CIM arrays. For further optimization by co-design ability, we can also optimize the N value for different high-performance (TD-P) and high-accuracy (TD-A) requirements.

### 3.3 KAN sparsity-aware weight mapping for $c_i'$

The parasitic resistance in BLs causes IR-drop, introducing computational errors during current-based summation in RRAM-ACIM's MAC operations, with consequent degradation of inference precision. While prior research [14] has attempted to address this challenge, existing solutions necessitate either supplementary circuit components or constraints on maximum array dimensions. The proposed KAN sparsity-aware weight mapping technique (KAN-SAM) circumvents these limitations by operating within existing hardware and algorithmic frameworks without requiring architectural modifications.

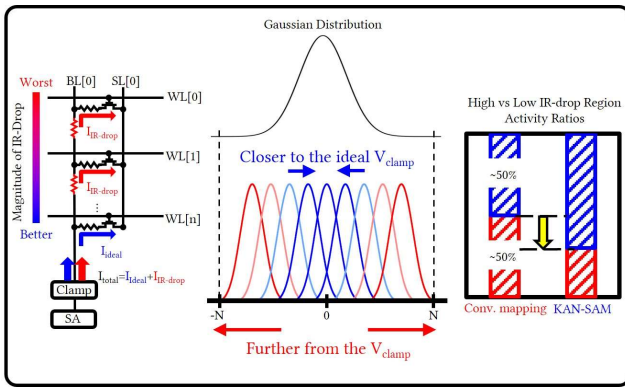
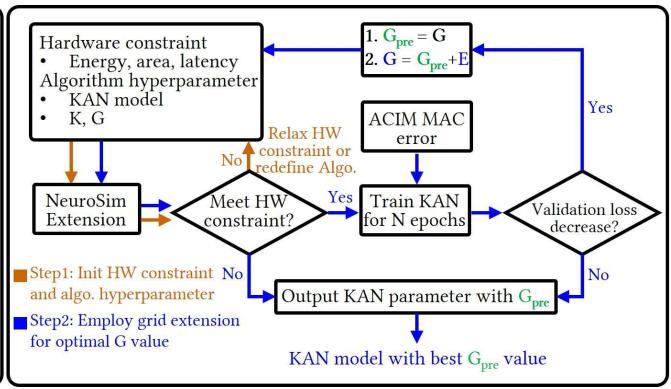

**Fig. 10.** KAN sparsity-aware weight mapping.   **Fig. 11.** KAN-NeuroSim hyperparameter optimization

The inherent characteristics of B(X) functions within KAN architectures dictate that only a subset of B(X) functions activate for any specific input value. In configurations where K=3, concurrent activation is limited to four B(X) functions. Through analysis of input probability distributions across data ranges, the B(X) functions exhibiting maximum activation likelihood can be identified, designated as B_H(X). Similarly, functions with minimal activation probability are denoted as B_L(X). The ci' coefficients associated with B_H(X) are strategically allocated to RRAM cells positioned proximate to BL clamping circuitry, preserving computational precision for frequently occurring inputs. In contrast, ci' coefficients linked to B_L(X) are assigned to cells at greater distances from clamping circuits, thus enhancing aggregate inference accuracy through probability-aware spatial optimization.

Diverse applications and model architectures demonstrate varying distributional characteristics. Fig. 10 exemplifies the KAN-SAM methodology utilizing a Gaussian distribution paradigm. As depicted in Fig. 10, within an input domain spanning [-N, N], the centrally located Bi(X) functions exhibit maximal activation probabilities, whereas boundary-positioned Bi(X) functions demonstrate minimal triggering likelihood. Therefore, the allocation of central ci' parameters (associated with B_H(X)) to RRAM cells adjacent to clamping circuits, coupled with the assignment of peripheral ci' parameters (linked to B_L(X)) to cells at increased distances, optimizes system-level inference precision. This mapping strategy remains applicable for input domains bounded by [0, N].

To ensure that KAN-SAM performs robustly across varying input distributions, we introduce Algorithm 1: KAN-SAM Strategy. Phase-A: In this phase we scan the training set once. For each basis Bi, record how often it fires (activation probability), its average magnitude when active, and how much it varies. Phase-B: Each trained coefficient becomes 8 binary slices stored on a fixed 8-column template in every row (most significant bit, MSB→least significant bit, LSB). During inference, slices are combined by shift-and-add, so we only optimize rows (distance), not columns. Phase C: For each coefficient, we build a score that favors three things: (i) bases that fire more often, (ii) bases that are stronger on average, and (iii) bases that are stable (low relative variability). Stability is derived from the coefficient of variation (standard deviation over mean) with a small numerical guard; then softly squashed to a 0~1 weight so unstable bases are de-emphasized without hard thresholds. A tunable mix combines "expected contribution" and "stability" into a single criticality score. Row mapping policy: Sort coefficients by the criticality score (high→low) and assign rows from nearest to farthest using a precomputed order. IR-drop grows with distance along the bit-line; giving the closest rows to the most impactful and stable coefficients reduces analog error where it matters most.

### 3.4 KAN-NeuroSim hyperparameter optimization framework

Section 3.1 presented the B(X) lookup optimization achieved through ASP-KAN-HAQ, which yields distinct LD values corresponding to different G parameter configurations. However, a systematic methodology for identifying optimal G parameters under hardware-imposed constraints remained unaddressed. Section 3.2 introduced the dual operational modes of TM-DVS-IG, namely the high-performance (TD-P) and high-accuracy (TD-A) configurations. Nonetheless, a comprehensive analytical framework for evaluating the respective impacts of TD-P and TD-A modes on system-level performance metrics has not been established, limiting the ability to guide mode selection based on application-specific requirements.

To overcome these constraints, the KAN-NeuroSim hyperparameter optimization framework is proposed, as illustrated in Fig. 11. The framework operates through a two-stage process. The initial stage, indicated by the brown pathway in Fig. 11, establishes hardware specifications (energy budget, silicon area, computational latency) alongside KAN architectural parameters (network topology, K, and G values). These specifications are subsequently processed through an extended NeuroSim implementation [17][28][29], which integrates ASP-KAN-HAQ and TM-DV-IG methodologies to derive energy consumption, area utilization, and latency metrics. When computed metrics violate hardware specifications, the

**Algorithm 1** KAN-SAM-Strategy

**Require:** Trained KAN coefficients $\{c'_i\}_{i=0}^{K+G-1}$; training set $D_{\text{train}}$; B-spline params $(K, G)$; crossbar with $R$ rows; precomputed row order RowOrder (nearest → farthest); hyperparameters $\alpha, \beta \in [0,1]$ with $\alpha + \beta = 1$, and $\varepsilon > 0$.

*Assumption: bit-sliced columns use a fixed 8-bit template across rows (MSB→LSB).*

**Phase A: Input-side statistics**
1: Initialize $\text{cnt}[i] \leftarrow 0, s_1[i] \leftarrow 0, s_2[i] \leftarrow 0$ for all $i$.
2: **for all** $x \in D_{\text{train}}$ **do**
3: $\quad A \leftarrow \text{active\_B\_indices}(x; K, G)$
4: $\quad$ **for all** $\{i \in A\}$ **do**
5: $\quad\quad b \leftarrow B_i(x)$ {Spline/LUT; $b \geq 0$}
6: $\quad\quad \text{cnt}[i] \leftarrow \text{cnt}[i] + 1; s_1[i] \leftarrow s_1[i] + b; s_2[i] \leftarrow s_2[i] + b^2$
7: $\quad$ **end for**
8: **end for**
9: **for** $i = 0$ to $K + G - 1$ **do**
10: $\quad p[i] \leftarrow \text{cnt}[i]/|D_{\text{train}}|$ {Activation probability}
11: $\quad \mu[i] \leftarrow \frac{s_1[i]}{\max(\text{cnt}[i],1)}$ {Arithmetic mean (for CV)}
12: $\quad \text{var}[i] \leftarrow \frac{s_2[i]}{\max(\text{cnt}[i],1)} - \left(\frac{s_1[i]}{\max(\text{cnt}[i],1)}\right)^2$
13: **end for**

**Phase B: Quantization and bit vectors**
14: **for** $i = 0$ to $K + G - 1$ **do**
15: $\quad b_i \leftarrow [b_{i,7}, \ldots, b_{i,0}] \in \{0,1\}^8$ {8-bit slices}
16: $\quad |c'_i|_Q \leftarrow \sum_{k=0}^{7} b_{i,k} 2^k$ {Digital magnitude}
17: **end for**

**Phase C: Coefficient criticality (CV-based stability)**
18: **for** $i = 0$ to $K + G - 1$ **do**
19: $\quad \sigma[i] \leftarrow \sqrt{\text{var}[i]}; \text{CV}[i] \leftarrow \frac{\sigma[i]}{\mu[i]+\varepsilon}$
20: $\quad S[i] \leftarrow \frac{1}{1+\text{CV}[i]} \in (0,1]$ {Monotone squashing of CV}
21: $\quad J[i] \leftarrow p[i] \cdot \mu[i] \cdot |c'_i|_Q$ {Expected contribution}
22: $\quad C_w[i] \leftarrow \alpha J[i] + \beta S[i] \cdot J[i]$
23: **end for**

**Row mapping policy**
24: Sort $i$ by $C\_w[i]$ ($high \rightarrow low$) to obtain $Q$; assign $Q[1], Q[2], \ldots$ to rows from nearest to farthest using RowOrder.

---

**Algorithm 2** Sensitivity-based Grid Assignment for KAN-NeuroSim

**Require:** Network architecture: $L$ (number of layers); Grid templates $G_{\text{high}}$, $G_{\text{med}}$, $G_{\text{low}}$; Training parameters: warmup_epochs

**Phase 1: Layer Sensitivity Profiling**
1: Initialize KAN model with uniform grid $G_{\text{init}}$
2: Train model for warmup_epochs
3: **for** $i = 1$ to $L$ **do**
4: $\quad$ Compute sensitivity: $S_i \leftarrow \mathbb{E}_{val}\left[\left(\frac{1}{M_i}\sum_{j=1}^{M_i}\left(\frac{\partial \mathcal{L}}{\partial c_{i,j}}\right)^2\right)\right]$
5: **end for**

**Phase 2: Sensitivity Classification and Grid Assignment**
6: Sort $S = [S_1, S_2, \ldots, S_L]$ in descending order
7: $\tau_{\text{high}} \leftarrow \text{percentile}(S, 67)$ {Top 33% are high sensitivity}
8: $\tau_{\text{low}} \leftarrow \text{percentile}(S, 33)$ {Bottom 33% are low sensitivity}
9: **for** $i = 1$ to $L$ **do**
10: $\quad$ **if** $S_i \geq \tau_{\text{high}}$ **then**
11: $\quad\quad G_i \leftarrow G_{\text{high}}$ {High sensitivity}
12: $\quad\quad class_i \leftarrow \text{``HIGH''}$
13: $\quad$ **else if** $S_i \geq \tau_{\text{low}}$ **then**
14: $\quad\quad G_i \leftarrow G_{\text{med}}$ {Medium sensitivity}
15: $\quad\quad class_i \leftarrow \text{``MEDIUM''}$
16: $\quad$ **else**
17: $\quad\quad G_i \leftarrow G_{\text{low}}$ {Low sensitivity}
18: $\quad\quad class_i \leftarrow \text{``LOW''}$
19: $\quad$ **end if**
20: **end for**
21: **return** $G^* = [G_1, G_2, \ldots, G_L]$

---

framework iteratively adjusts either the constraint parameters or KAN hyperparameters until compliance is achieved. Following successful constraint satisfaction, the secondary stage employs the grid extension methodology established in the original KAN literature to enhance computational accuracy. Throughout the training procedure, grid expansion occurs at N-epoch intervals. The parameter G undergoes incremental augmentation by a user-specified value E, contingent upon sustained validation loss reduction and compliance with hardware resource boundaries as determined through NeuroSim evaluation. When these criteria are not satisfied, the grid extension process is terminated, with the system reverting to the preceding $G_{\text{pre}}$ configuration. The framework incorporates RRAM non-ideality factors, particularly partial sum error characteristics, derived from statistical measurements conducted on TSMC 22nm RRAM-ACIM prototype chip. This integration guarantees that the resulting KAN hyperparameters deliver optimized hardware performance and inference accuracy when deployed on RRAM-ACIM systems.

To enable users to achieve better performance under limited hardware constraints, we introduce Algorithm 2: Sensitivity-based Grid Assignment for KAN-NeuroSim. This strategy allows users to allocate larger G values to regions of the network that exhibit higher sensitivity in order to preserve accuracy, while assigning smaller G values to less sensitive regions to reduce hardware requirements. The algorithm operates in two phases. First, during a warmup training period, we profile each layer's sensitivity by computing the gradient. This sensitivity metric quantifies the degree of sensitivity for each region. In the second phase, layers are classified into three sensitivity tiers based on percentile thresholds. High sensitivity layers (top 33%) are assigned $G_{\text{high}}$, as they require finer grid resolution for accurate feature extraction. Medium sensitivity layers (middle 34%) receive $G_{\text{med}}$, while low sensitivity layers (bottom 33%) operate efficiently with $G_{\text{low}}$. This heterogeneous assignment ensures that computational resources are allocated where they provide the most benefit. Following the sensitivity-based grid assignment, KAN-NeuroSim implements a two-step optimization process as shown in Fig. 11. It should be noted that users may autonomously determine the granularity of grid templates. Three grid resolution levels—namely $G_{\text{high}}$, $G_{\text{med}}$, and $G_{\text{low}}$—are utilized here as examples, and variable grid resolutions can be assigned within the same layer based on sensitivity requirements.

## 4 EVALUATION RESULTS

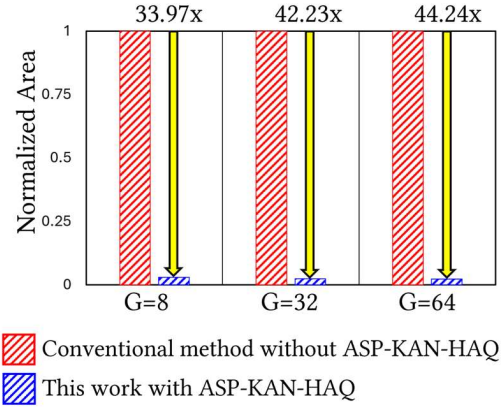

**Fig. 12.** Comparison of Normalized Area between proposed ASP-KAN-HAQ and conventional method based on Post-Training Quantization [29] using NVIDIA's TensorRT framework.

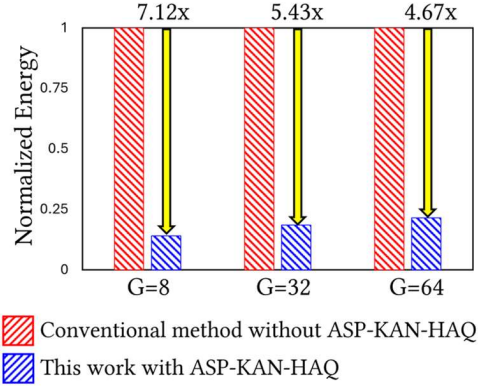

**Fig. 13.** Comparison of Normalized Energy Consumption between proposed ASP-KAN-HAQ and conventional method based on Post-Training Quantization [29] using NVIDIA's TensorRT framework.

A. ASP-KAN-HAQ

We utilize a large-scale task [23] for evaluating ASP-KAN-HAQ, employing the CF-KAN architecture—an encoder-decoder framework based on KAN designed for recommendation systems. As outlined in Section 3.1, the parameter G serves as a key factor in ASP-KAN-HAQ. To systematically evaluate our method's scalability across more complex KAN architectures with arbitrary G values, we progressively increased G using the grid extension approach, a methodology established by the original KAN paper authors. Our experimental setup utilized TSMC 22nm technology node parameters. The evaluation environment incorporated detailed circuit-level simulations using SPICE for accurate power and timing analysis, while area estimations were derived from synthesized netlists. We evaluated ASP-KAN-HAQ against conventional quantization approaches [15][16] with respect to energy and area metrics at the 22nm technology node. For this study, Post-Training Quantization [29] implemented via NVIDIA's TensorRT framework served as our comparison baseline. To isolate variables, we concentrated our analysis on the hardware pathway from the input X, through LUT-based

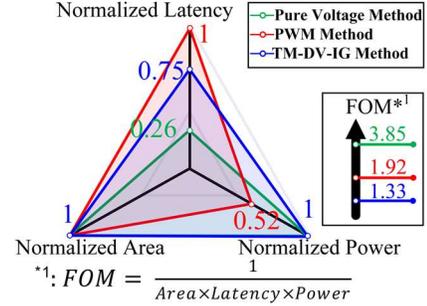

**Fig.14.** WL input methods performance comparison with SPICE simulation at 22 nm for N=1 2-bit vector input scheme.

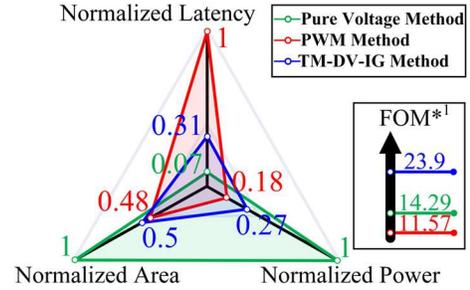

**Fig.15.** WL input methods performance comparison with SPICE simulation at 22 nm for N=2 4-bit vector input scheme.

B(X) value retrieval, to its delivery to the input generator. The evaluation encompasses various precision levels ranging from 4-bit to 8-bit quantization, with 8-bit selected as the optimal trade-off between accuracy and hardware efficiency. Fig. 12 and Fig. 13 illustrate the efficacy of our approach. With G scaling from 8 to 64, our method exhibits substantial advantages over conventional techniques, delivering an average area reduction of 40.14× and an average energy reduction of 5.74×. Specifically, at G=8, our method achieves 33.97× area reduction and 7.12× energy savings, while at G=64, these improvements scale to 44.24× and 4.67× respectively. The energy efficiency gains are primarily attributed to simplify decoder structure and the LUT reduction achieved by the proposed SH-LUT architecture. This can be attributed to inherent constraints in conventional quantization approaches, wherein non-zero offsets between quantization and knot grids hinder the ability to share LUTs across different B(X) values. In contrast, our approach enables all B(X) values to utilize a unified LUT while separating local and global information, thereby reducing TG-MUX and decoder areas and maintaining KAN scalability at the edge through ASP-KAN-HAQ.

B. N:1 Time Modulation Dynamic Voltage input generator

To quantitatively evaluate the benefits of the proposed N:1 TM-DV-IG for KAN accelerator implementation, we compared its performance against conventional pure voltage and pure PWM input schemes across multiple WL resolutions, as shown in Fig. 14–17. Benchmark simulations were performed for N=1-

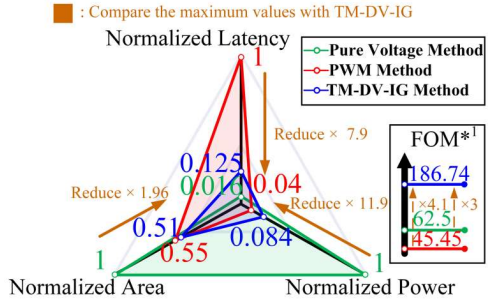

**Fig.16.** WL input methods performance comparison with SPICE simulation at 22 nm for N=3 3-bit vector input scheme.

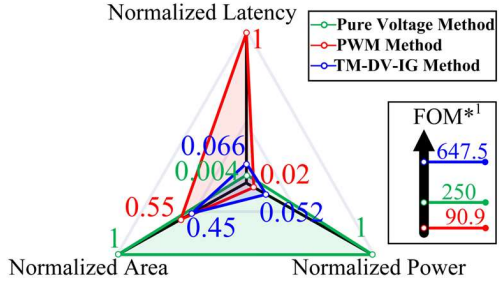

**Fig.17.** WL input methods performance comparison with SPICE simulation at 22 nm for N=4 8-bit vector input scheme.

4, which means 2-, 4-, 6-, and 8-bit input vectors, corresponding to $2^2$, $2^4$, $2^6$, and $2^8$ distinct WL pulses for BL sampling, respectively. For fair comparison, the unit pulse width was assigned identically to all three methods, as illustrated in Fig. 8(b) and (c). The evaluation was conducted in a 22 nm technology node, and all circuit modules were validated at the transistor level using SPICE. The results reveal distinct trade-offs across the three methods. In the 2-bit input case (Fig. 14), the pure voltage scheme achieves the best figure-of-merit (FOM) due to its minimal latency, while PWM provides superior power efficiency. The TM-DV-IG in this low-resolution case exhibits the lowest FOM, primarily due to its relatively redundant circuit structure. However, when the input resolution increases (N>1, i.e., 4-, 6-, and 8-bit vectors), the advantages of TM-DV-IG become increasingly evident. For the 6-bit scheme (N=3), although the pure voltage method still achieves the lowest latency, it requires a high-resolution DAC, which reduces noise margin and incurs significant static power consumption. Specifically, it suffers from a 1.96× area overhead and an 11.9× power overhead compared with TM-DV-IG. The pure PWM method shows the poorest performance, exhibiting an 8× latency overhead and a 1.07× area overhead due to the long delay chain requirement. In contrast, TM-DV-IG, by combining voltage and timing modulation, avoids the noise margin limitations of the high-bit DAC and the excessive timing overhead of PWM, thereby achieving superior overall efficiency. When all three metrics—area, power, and latency—are jointly considered, TM-DV-IG delivers the highest FOM once N>1. In the 6-bit configuration as shown in Fig.16, it achieves 3× improvement over pure voltage and 4.1× improvement over pure PWM. In the 8-bit scheme (N=4), an 8-bit input vector is directly applied to the WL as shown in Fig.17, and TM-DV-IG continues to demonstrate significant FOM improvements over both conventional approaches. These results confirm that the proposed TM-DV-IG effectively balances latency, area, and power, making it a scalable and efficient enabler for KAN algorithm implementation on RRAM-based CIM architectures.

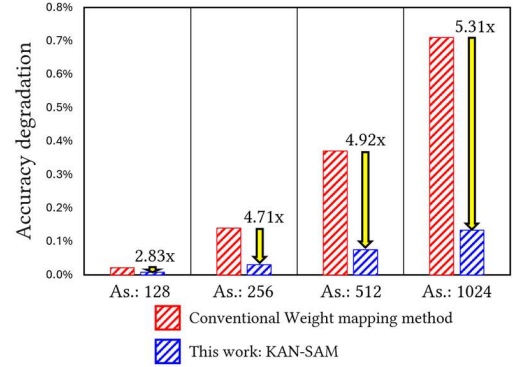

**Fig. 18.** Comparison of accuracy degradation from KAN software baseline across different RRAM array sizes (As.). The statistics of measured RRAM-ACIM chips [13] are used.

| Metrics | [27] | CF-KAN-1 | CF-KAN-2 |
|---|---|---|---|
| #Param | 78 B | 39 MB | 63 MB |
| Task Scale | Tiny | Large | Large |
| Area (mm²) | 0.0034225 | 97.76 | 142.24 |
| Energy (nJ) | - | 289.6 | 645.9 |
| Power (W) | 0.001547 | 0.079 | 0.146 |
| Latency (ns) | - | 3648 | 4416 |
| Accuracy degradation | 2.02% | 0.23% | 0.11% |
| Technology | 28nm | 22nm | 22nm |

**Fig. 19.** Comparison of proposed KAN accelerator with previous work across small-scale and large-scale computational tasks.

### C. KAN sparsity-aware weight mapping

We estimated the IR-drop issue and evaluated the proposed KAN-SAM architecture. The IR-drop phenomenon in RRAM-ACIM systems occurs when multiple cells along a bit-line are activated simultaneously, causing voltage degradation that directly impacts computation accuracy. Firstly, we refer to TSMC's 22 nm RRAM-ACIM chip measurement results [13] of the single BL IR drop effect in array sizes ranging from 128 to 1024. These measurements provide empirical validation of the voltage drop characteristics under different array configurations, establishing a reliable foundation for our error modeling.

Secondly, MAC error rates induced by IR-drop were extracted from TSMC's 22 nm RRAM-ACIM chips, which were subsequently used to train four CF-KAN models in PyTorch, employing G values of 7, 15, 30, and 60 that correspond to array dimensions of 128, 256, 512, and 1024, respectively. The choice of these specific G values ensures comprehensive coverage of practical RRAM array dimensions.

The baseline approach applied uniform mapping of different $c_i'$ values to RRAM-ACIM without accounting for $B_i(X)$ activation probabilities. Through the integration of extracted MAC error rates and variable $B_i(X)$ activation probabilities, we evaluated the influence of KAN-SAM on accuracy performance. Our sparsity-aware mapping strategy strategically places weights with higher activation probabilities in array positions less susceptible to IR-drop effects, effectively minimizing the overall computational error. Fig. 18 illustrates that with array dimensions increasing from 128 to 1024, KAN-SAM achieves accuracy enhancements ranging from 2.83× to 5.31×. This progressive improvement underscores KAN-SAM's capability in improving the scalability of RRAM-ACIM systems.

D. KAN-NeuroSim hyperparameter optimization framework

We employed KAN-NeuroSim within a PyTorch environment to optimize the G value for KAN architecture under various hardware constraints for large-scale recommendation system tasks, utilizing the Anime dataset for our analysis. In this study, we explored two architectures respectively targeting high performance mode (CF-KAN-1) and high accuracy mode (CF-KAN-2). When searching for optimal parameters for high performance, we implemented Algorithm 2, a Sensitivity-based Grid Assignment strategy for KAN-NeuroSim, ensuring minimal Grid utilization in non-sensitive regions to enhance hardware performance, while allocating additional Grid resources in sensitive regions to prevent significant accuracy degradation. Furthermore, we deployed TM-DV-IG's TD-P mode in non-sensitive regions to reduce latency and energy consumption, while employing TD-A mode in sensitive regions to maintain high accuracy when executing complex large-scale tasks. Regarding the high accuracy mode (CF-KAN-2), Algorithm 2 was disabled to ensure optimal accuracy, with the entire network utilizing Ghigh and operating in TD-A mode to guarantee maximum accuracy. Fig. 19 demonstrates the performance of this work, showing energy consumption of 289.6 nJ and 645.9 nJ for large-scale tasks with parameter counts of 39 MB and 63 MB, respectively, with corresponding latencies of 3648 ns and 4416 ns, while accuracy degradation remained minimal at only 0.23% and 0.11%.

Compared to previous work operating exclusively on tiny-scale tasks, CF-KAN-1 and CF-KAN-2 have parameter counts 500K and 807K times larger than [27], yet the area increased by only 28K and 41K times, respectively. This is attributed to our efficient ASP-KAN-HAQ, which significantly reduces the hardware resources required for LUTs. On the other hand, power consumption increased by 51× and 94×, respectively, due to: 1) ASP-KAN-HAQ's reduction in energy consumption, 2) our highly integrated system leveraging the high-performance parallel computing capabilities of RRAM-ACIM, and 3) TM-DV-IG achieving optimal trade-offs among accuracy, energy consumption, latency, and area. Therefore, this work maintains high efficiency even in large-scale applications with extremely high parameter counts.

## 5 Conclusion

This work introduces an innovative hardware acceleration methodology for KAN through algorithm-hardware co-design. The proposed algorithmic and circuit-level innovations effectively minimize hardware overhead, power consumption, and maintain inference accuracy for resource-constrained edge computing applications. To the best of our knowledge, this work represents the first validation of large-scale tasks on KAN accelerators. Evaluation results demonstrate that, compared to previous work operating on tiny tasks, despite the parameter count for large-scale tasks in this work increasing by 500K× to 807K×, the area overhead increases by only 28K× to 41K×, while power consumption increases by merely 51× to 97×, with accuracy degradation remaining minimal at 0.11% to 0.23%, thereby showcasing the exceptional scaling capability of our proposed architecture.

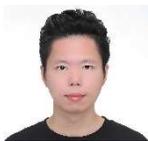
Wei-Hsing Huang received the B.S. degree in electrical engineering from the National Chung Cheng University, Chiayi, Taiwan, in 2017, and the M.S. degree in electrical engineering and computer science from the National Tsing Hua University, Hsinchu, Taiwan, in 2019. He is currently a Research Assistant in electrical and computer engineering with Georgia Institute of Technology, Atlanta, GA, USA. His current research interests include deep learning algorithms and algorithm-hardware co-design for deep learning.

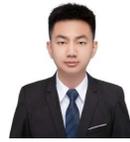
Jianwei Jia received the B.S. degree in Microelectronics Science and Engineering from Nankai University, Tianjin, China, in 2021, and the M.S. degree in Electrical and Computer Engineering from the University of Michigan, Ann Arbor, MI, USA, in 2023. He is currently pursuing the Ph.D. degree at the Georgia Institute of Technology, Atlanta, GA, USA.

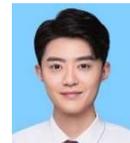
Yuyao Kong received the B.S. degree from Nanjing Tech University, Nanjing, China, in 2015, the M.S. degree from the University of Southampton, Southampton, U.K., in 2016, and the Ph.D. degree from the School of Electronic Science and Engineering, Southeast University, Nanjing, China, in 2023. He is currently a Postdoctoral Fellow with the Laboratory for Emerging Devices and Circuits, Georgia Institute of Technology, advised by Prof. Shimeng Yu. His research interests include compute-in-memory (CIM)-based algorithm-hardware co-design targeting AI processors and probabilistic computing, as well as low-voltage SRAM and other energy-efficient circuit designs.

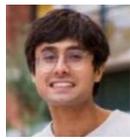
Faaiq Waqar received a B.S. degree in computer science and electrical & computer engineering from Oregon State University, Corvallis, OR, in 2022. He is currently pursuing a Ph.D. in electrical & computer engineering from the Georgia Institute of Technology, Atlanta, GA. Prior to joining Georgia Tech, he worked as a hardware engineer for Microsoft's Silicon Engineering Solutions team. He was the recipient of the NSF Graduate Research Fellowship and the Georgia Tech President's Fellowship in 2023. His current research interests pertain to the modeling and metrology of emerging amorphous oxide semiconductor and ferroelectric devices for applications in neuromorphic, reconfigurable, and high-performance computational systems.

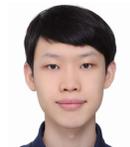
Tai-Hao Wen (Member, IEEE) received the B.S. degree in electrical engineering from National Tsing Hua University, Hsinchu, Taiwan, in 2020, and the Ph.D. degree in electrical engineering from National Tsing Hua University, Hsinchu, Taiwan, in 2024. He is currently a Postdoctoral Research Fellow with the Department of Electrical and Computer Engineering, University of Michigan, Ann Arbor, USA. His research interests include memory circuit design and compute-in-memory for emerging nonvolatile memories, as well as hardware-efficient system design.

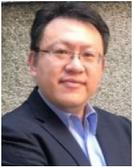
Meng-Fan Chang received the M.S. degree from The Pennsylvania State University, State College, PA, USA, and the Ph.D. from the National Chiao Tung University, Hsinchu, Taiwan. Prior to 2006, he worked in the industry for over ten years. This included the design of memory compilers (Mentor Graphics, Wilsonville, OR, USA, from 1996 to 1997) and the design of embedded SRAM and Flash macros (Design Service Division, TSMC, Hsinchu, From 1997 to 2001). In 2001, he co-founded IPLib, Hsinchu, where he developed embedded SRAM and ROM compilers, flash macros, and flat-cell ROM products, until 2006. He is currently a Distinguished Professor at the National Tsing Hua University (NTHU) and the Director of Corporate Research, TSMC. His research interests include circuit design for volatile and nonvolatile memory, ultralow-voltage systems, 3-D memory, circuitdevice interactions, spintronic circuits, memristor logics for neuromorphic computing, and computing-in-memory for artificial intelligence.

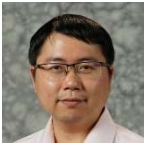
Shimeng Yu (Fellow, IEEE) is a full professor of electrical and computer engineering at Georgia Institute of Technology, where he holds the Dean's Professorship. He received the B.S. degree in microelectronics from Peking University in 2009, and the M.S. degree and Ph.D. degree in electrical engineering from Stanford University in 2011 and 2013, respectively. From 2013 to 2018, he was an assistant professor at Arizona State University. He is elevated for the IEEE Fellow for contributions to non-volatile memories and in-memory computing. His general research interests are semiconductor devices and integrated circuits for energy-efficient computing systems. His expertise is on the emerging non-volatile memories for AI hardware and 3D integration. Prof. Yu's 400+ journal/conference publications received more than 30,000 citations with H-index 82. He is the theme lead of two SRC/DARPA JUMP 2.0 centers on intelligent memory/storage and heterogeneous/monolithic 3D integration.